\documentclass[aps,prl,twocolumn,showpacs,superscriptaddress,lengthcheck]{revtex4}
\usepackage{eurosym}
\usepackage{graphicx}
\usepackage{amssymb}
\usepackage{amsmath}
\usepackage{color}

\begin{document}

\title{Damping of dHvA oscillations and vortex-lattice disorder in the
peak-effect region of strong type-II superconductors}
\author{A. Maniv}
\affiliation{NRCN, P.O. Box 9001, Beer Sheva, 84190, Israel}
\author{T. Maniv}
\altaffiliation{e-mail:maniv@tx.technion.ac.il}
\affiliation{Schulich Faculty of Chemistry, Technion-Israel Institute of Technology,
Haifa 32000, Israel}
\author{V. Zhuravlev}
\affiliation{Schulich Faculty of Chemistry, Technion-Israel Institute of Technology,
Haifa 32000, Israel}
\author{B. Bergk}
\affiliation{Hochfeld-Magnetlabor Dresden (HLD), Forschungszentrum Dresden-Rossendorf,
D-01314 Dresden, Germany}
\author{J. Wosnitza}
\affiliation{Hochfeld-Magnetlabor Dresden (HLD), Forschungszentrum Dresden-Rossendorf,
D-01314 Dresden, Germany}
\author{A. K\"ohler}
\affiliation{Leibniz-Institut f\"ur Festk\"orper- und Werkstoffforschung Dresden, D-01171
Dresden, Germany}
\author{G. Behr}
\affiliation{Leibniz-Institut f\"ur Festk\"orper- und Werkstoffforschung Dresden, D-01171
Dresden, Germany}
\author{P. C. Canfield}
\affiliation{Ames Laboratory and Department of Physics, Iowa State University, Ames, Iowa
50011, USA}
\author{J. E. Sonier}
\affiliation{Department of Physics, Simon Fraser University, Burnaby, British Colombia
V6T 1Z1, Canada}
\affiliation{Canadian Institute for Advanced Research, Toronto, Ontario, Canada}
\date{\today }

\begin{abstract}
The phenomenon of magnetic quantum oscillations in the superconducting state
poses several questions that still defy satisfactory answers. A key
controversial issue concerns the additional damping observed in the vortex
state. Here, we show results of $\mu $SR, dHvA, and SQUID magnetization
measurements on borocarbide superconductors, indicating that a sharp drop
observed in the dHvA amplitude just below $H_{c2}$ is correlated with
enhanced disorder of the vortex lattice in the peak-effect region, which
significantly enhances quasiparticle scattering by the pair potential.
\end{abstract}

\pacs{74.25.Ha, 74.25.Uv, 74.70.Dd}
\maketitle

The recent observations of Shubnikov--de Haas (SdH) and de Haas--van Alphen
(dHvA) oscillations in high-temperature copper-oxide superconductors \cite%
{Jaudet08} have drawn increasing attention to magnetic quantum oscillations
(MQO) as a key technique for investigating the superconducting (SC) states
of these materials at high magnetic fields. However, the usefulness of these
studies depends crucially on the existence of a reliable quantitative theory
for MQO in the vortex state of strong type-II superconductors, a theory
which does not exist today even for conventional type-II superconductors 
\cite{Maniv01}. The demanding situation involved in developing such a theory
may be illustrated by the extensive investigations performed on the
nonmagnetic borocarbide superconductors YNi$_{2}$B$_{2}$C, and LuNi$_{2}$B$%
_{2}$C \cite{Terashima97,Isshiki08,Maniv06,Bergk09}. Terashima \textit{et al.%
}\ \cite{Terashima97} applied the field-modulation technique to YNi$_{2}$B$%
_{2}$C, finding a strong suppression of the dHvA amplitude just below the
entrance to the SC state, followed by a recovery of the signal at slightly
lower fields, and a very smooth additional damping over a large field range
below the SC transition. The small region of strong additional damping was
found to correlate with that of a significant peak observed in the measured
magnetization (the so-called peak-effect, PE). Remarkable persistence of the
dHvA signal deep in the vortex state of LuNi$_{2}$B$_{2}$C has also been
reported by Isshiki \textit{et al.}\ \cite{Isshiki08}.

Similarly, large additional damping of the dHvA signal, coinciding with the
onset of the PE, was recently observed in YNi$_{2}$B$_{2}$C by employing the
cantilever torque technique \cite{Maniv06}, with the signal persisting at
significantly lower fields (down to 3 T). Finally, in a series of dHvA
measurements, carried out very recently on LuNi$_{2}$B$_{2}$C by the
field-modulation technique \cite{Bergk09}, the salient features reported for
YNi$_{2}$B$_{2}$C \cite{Terashima97, Maniv06} have been confirmed (Fig.\ \ref%
{Fig.1}).

The current theoretical approaches to MQO in the vortex state \cite{Maniv01}
do not provide a consistent predictive framework for a quantitative
interpretation of the intriguing experimental results\textbf{.} The
mean-field theories based on a detailed exposition of the quasi-particle
excitations obtained by solving the corresponding Bogoliubov--de Gennes
equations for an ordered vortex lattice \cite{Dukan94,Norman95,Kita02},
provide deep insight into fine features of the Landau band structure, but
lose their transparency very quickly and become heavily numerical at early
stages of the analysis. A simple formula for the additional damping \cite%
{Maki91}, used frequently in the literature, has been shown to be
essentially valid only in the limiting case of a random vortex distribution 
\cite{Maniv01}. The resulting additional damping rate of MQO seriously
overestimates the rate calculated numerically (see Fig.\ 8 in Ref.\ \cite%
{Kita02}) for an Abrikosov vortex lattice.

This may indicate that enhanced quasi-particle scattering by the SC
pair-potential due to strong disorder of the vortex lattice in the PE region
is responsible for the enhanced additional damping reported in Refs. \cite%
{Terashima97,Maniv06,Bergk09}. However, as proposed in Ref.\ \cite%
{Terashima97}, an increased "phase-smearing" effect, i.e., broadening of
Landau levels by the inhomogeneity of the internal magnetic field associated
with displaced flux lines by random pinning centers, might also be
responsible for the enhanced damping in the PE region.

In the present paper it is shown, by means of $\mu $SR and complementary
SQUID magnetization measurements, that phase smearing is much too small to
account for the enhanced additional damping. It is also found that strong
disorder of the vortex lattice, revealed by characteristic change of the $%
\mu $SR line shape, is correlated with the strong suppression of the dHvA
signal in the PE region, whereas the establishment of a well-ordered vortex
lattice well below the PE region closely follows the weak additional damping
of the dHvA signal, observed in this broad field-range. Both findings
strongly support a scenario whereby quasiparticle scattering is enhanced by
the disorder of the vortex lattice.

\begin{figure}[tbp]
\includegraphics[scale=.5]{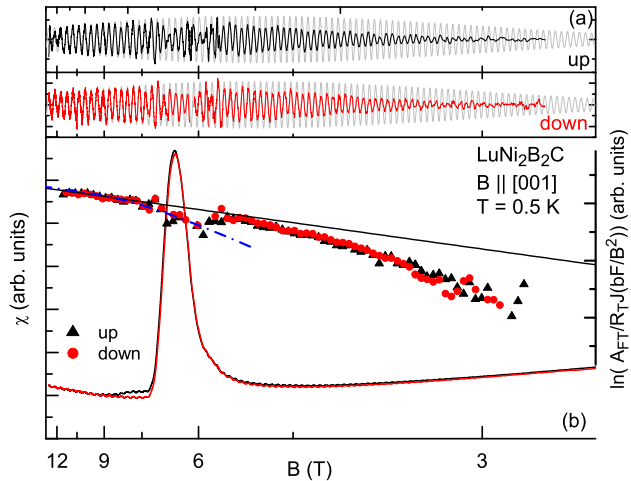}
\caption{(color online) (a) dHvA oscillation signals, for upward and
downward field-sweeps respectively, measured on LuNi$_{2}$B$_{2}$C after
background subtraction, and (b) the corresponding Dingle plots (triangles
and circles respectively).\emph{\ }The error arising from the background
subtraction analysis do not exceed the diamater of the data points. The gray
oscillatory lines in (a) represent the extrapolated dHvA signal, based on
the normal state Lifshitz-Kosevich formula, whereas the solid straight line
in (b) is the corresponding Dingle plot. The total magnetization for both
field-sweeps (solid curves) are also shown in (b). The dash-dotted line in
(b) represents the result of a calculation based on the random vortex
distribution model with the zero field order parameter $\Delta _{0}=4\ $meV
and mean-field $H_{c2}=8$ T. Measurements were done at 0.5 K.}
\label{Fig.1}
\end{figure}

High-quality single crystals of LuNi$_{2}$B$_{2}$C\emph{\ }and YNi$_{2}$B$%
_{2}$C were prepared by the flux-growth technique at Ames Laboratory, USA 
\cite{Can98}, and by the zone-melting method at the IFW Dresden, Germany 
\cite{Behr99}, respectively. The SQUID magnetization measurements were
performed at the Dresden High Magnetic Field Laboratory, Germany, while the $%
\mu $SR experiments were carried out at TRIUMF, Vancouver, Canada. The dHvA
data exploited in the comparative analysis was extracted from previouse
measurements employing the field-modulation technique on LuNi$_{2}$B$_{2}$C
at $T=0.5$~K \cite{Bergk09}. A detailed discussion of their dHvA signal in
the normal state and the electronic band structure is given in Ref. \cite%
{Bergk08}. Note, that due to technical reasons the $\mu $SR measurements
were carried out on samples similar (but not identical) to the ones where
the dHvA effect was measured. Furthermore, it is not currently feasible to
carry out $\mu $SR measurements below 2~K in fields above 5~T. Thus, in
order to allow cross-correlation between dHvA and $\mu $SR measurements, the
magnetization was also measured using a SQUID magnetometer on the same
samples and at the same temperatures as employed in the $\mu $SR experiments.

Transverse-field (TF) $\mu $SR measurements up to 7~T were carried out on
the M15 muon beam line at TRIUMF using the HiTime spectrometer, which
consists of muon and positron detectors contained within a standard He-flow
cryostat. The external field was directed parallel to the c-axis of each
crystal. A fast Fourier transform (FFT) of the TF-$\mu $SR signal closely
resembles the internal magnetic-field distribution $P(B)$ \cite{S13}. The
measurements were typically done by cooling the sample in a fixed field to a
temperature between 2 and 3~K, and then measuring the field dependence of
the TF-$\mu $SR signal. Specifically, measurements were performed on the LuNi%
$_{2}$B$_{2}$C single crystal after field-cooling in 3 and 7~T, and on YNi$%
_{2}$B$_{2}$C after field-cooling in 0.5 and 3~T. Several
temperature-dependent measurements were also made. In addition, measurements
were performed on each sample above $T_{c}$ at 20~K to visualize the
broadening of the TF-$\mu $SR line shape by nuclear moments and the field
inhomogeneity of the external magnet.

\begin{figure}[tbp]
\begin{center}
$%
\begin{array}{c}
\hspace*{.6cm} \includegraphics[scale=.75]{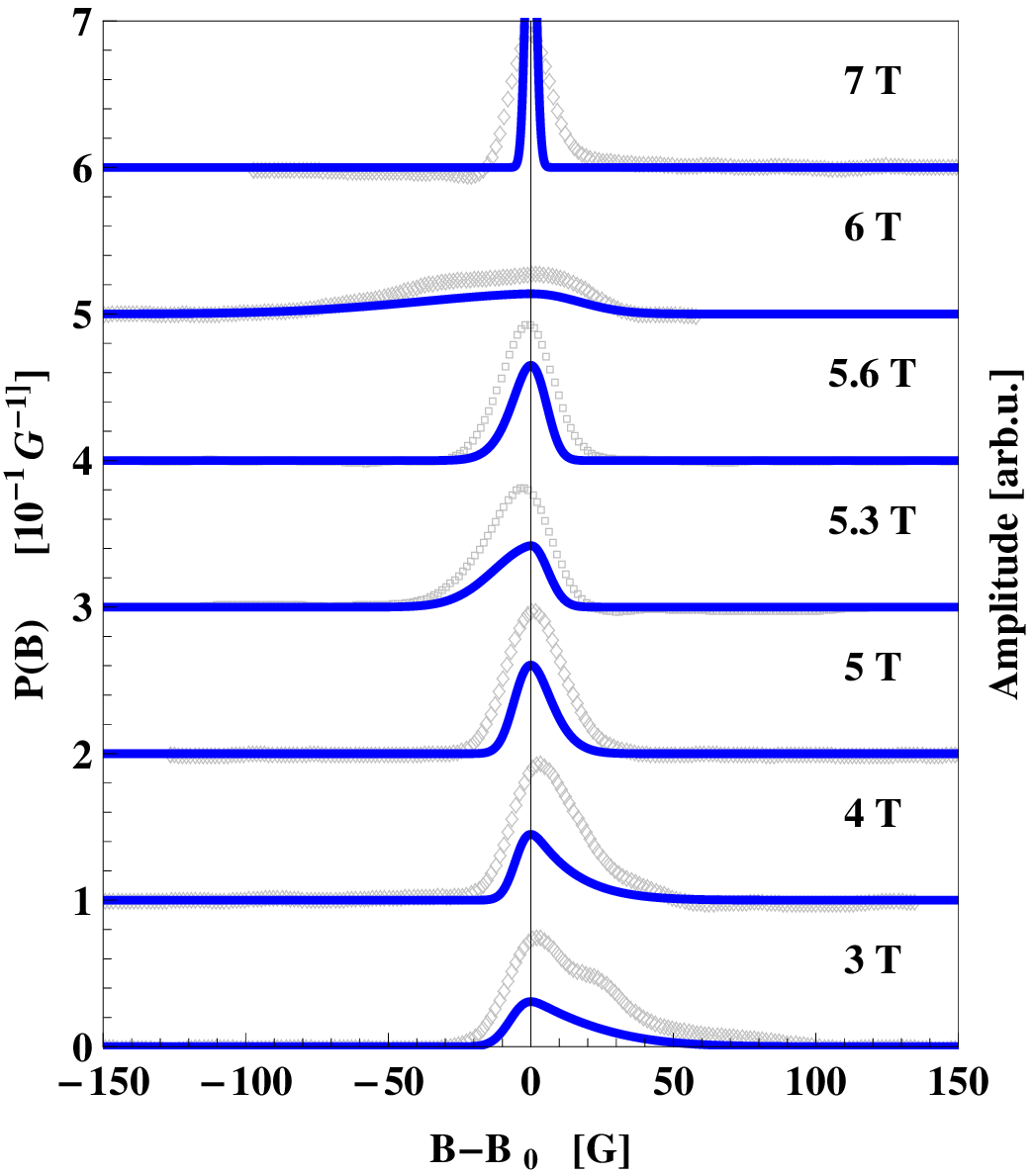} \\ 
\includegraphics[width=7.5cm]{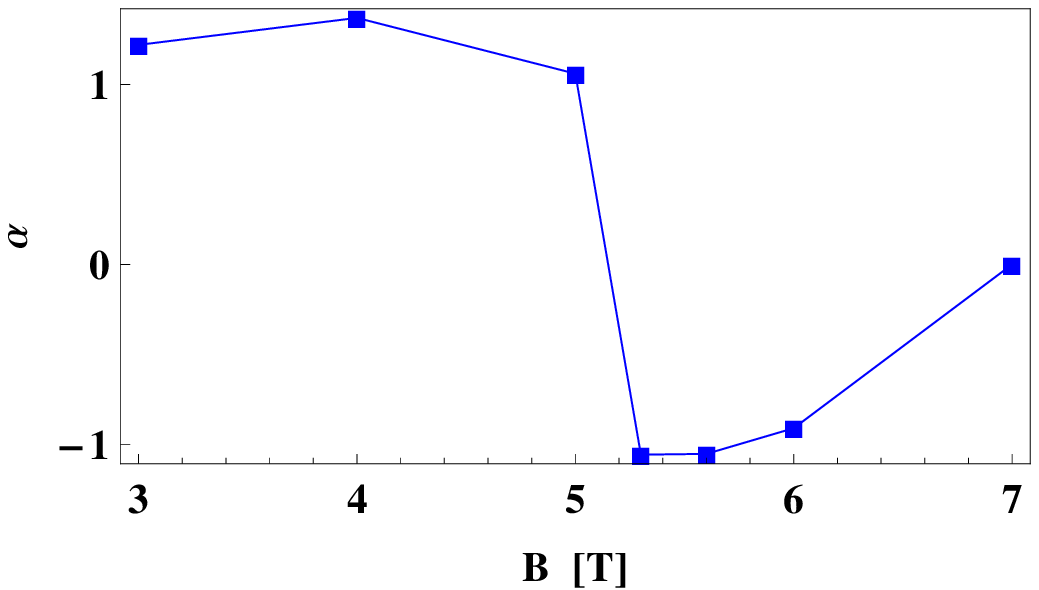}%
\end{array}%
$%
\end{center}
\caption{(color online) (upper panel) Probability field-distribution lines, $%
P(B)$, for LuNi$_2$B$_2$C (solid lines) at different external magnetic
fields, derived by deconvoluting the FFTs (open symbols) of the $\protect\mu$%
SR signals. Measurements at $H=$ 3, 4, 5, and 6 T were done after
field-cooling to 2.3~K at 3~T, whereas at $H=$ 5.3 and 5.6~T were performed
after degaussing at 2.7~K. The lines were displaced vertically for clarity
and horizontally for comparison. The reference data used in the
deconvolution were measured on the same sample above $T_{c}$ at $T=20$~K.
The shoulder seen on the 3~T FFT curve around 25~G is due to muons which
missed the SC sample. (lower panel) Field-dependence of the skewness
parameter, $\protect\alpha$ (see text), as calculated from the field
distributions shown in the upper panel. The line connecting the data points
is a guide to the eyes.}
\label{Fig.2}
\end{figure}

A typical result from a series of systematic dHvA measurements on the
flux-grown LuNi$_{2}$B$_{2}$C crystal \cite{Bergk09} is shown in Fig.\ \ref%
{Fig.1} for field parallel to the c-axis. The drop seen in the Dingle plot
of the dHvA signal (originating from the spherical Fermi surface \cite%
{Bergk08}) just below the SC transition is nicely correlated with the PE
seen in the magnetization envelope. This feature and the weak additional
attenuation of the dHvA signal seen below the PE region, essentially agree
with the results reported for YNi$_{2}$B$_{2}$C in Refs.\ \cite{Terashima97,
Maniv06}. One should note that the extraction of the dHvA signal at the PE
region is somewhat ambiguous, since a reliable determination of the
non-oscillating PE background signal is challenging. In our analysis the
background signal was subtracted by fitting suitable polynomials to
different sections of the raw-data signal. The Dingle data was generated
using step-by-step Fourier transforms over 3 oscillations. The corresponding
error bar of the amplitude in the PE range is about 15 percent, ensuring our
observation of a significant additional damping there. For other field
orientations somewhat different damping of the dHvA signals at the PE region
was found (see e.g. Fig. 7 in Ref. \cite{Bergk09}).

\begin{figure}[tbp]
\includegraphics[scale=.3]{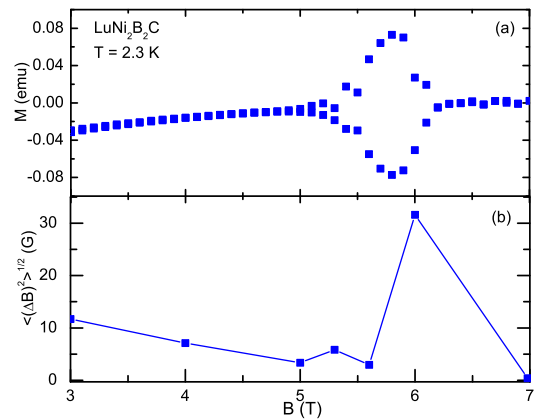}
\caption{(color online) Field dependence of the longitudinal magnetic moment
LuNi$_2$B$_2$C around the PE region obtained by SQUID magnetization
measurements at $T=2.3$ K (upper panel), and the corresponding field
dependence of the $\protect\mu$SR field distribution width (lower panel)
calculated from the data shown in Fig.\ \protect\ref{Fig.2}. The line
connecting the data points is a guide to the eyes.}
\label{Fig.3}
\end{figure}

The upper panel of Fig. \ref{Fig.2} shows FFTs of the measured $\mu $SR
signals for LuNi$_{2}$B$_{2}$C, after field cooling to $2.3$ K in $3$ T, and
the corresponding probability field distribution, $P(B)$, obtained by
deconvoluting each FFT curve with respect to the (practically Gaussian)
reference signal. The onset of a large broadening of $P(B)$ and its reversed
skewness in a small field range around $6$ T are apparent. The dramatic
skewness reversal of the line shape is illustrated in the lower panel of
Fig. \ref{Fig.2}, where the skewness parameter $\alpha =\left\langle \Delta
B^{3}\right\rangle ^{1/3}/\left\langle \Delta B^{2}\right\rangle ^{1/2}$,
with $\Delta B=B-\left\langle B\right\rangle $ \cite{S13} is plotted for the
various distribution functions. Note, that a negative $\alpha $ is due to
the presence of short-range triplet correlation in the absence of long-range
order, characterizing a vortex-glass phase \cite{Menon}. Thus, the onset of
negative $\alpha $ (around 5 T in Fig. \ref{Fig.2}) just below the PE region
(Fig. \ref{Fig.3}) indicates that the vortex lattice is disordered in the
entire PE region. \ Remarkably, the sharp change of $\alpha $ is seen to
correlate with the appearance of the PE and the additional damping of the
dHvA oscillation shown in Fig.\ \ref{Fig.1}. The positive values of $\alpha $
near unity in a broad field range below the PE reflect the occurrence of a
well-ordered vortex lattice in this region, which is seen to correlate with
the weak additional damping of the dHvA signal. One should be aware,
however, of the kinetic nature of the PE, which is reflected by the history
dependence of the measured $\mu $SR data. For example, field-cooling in $3$%
~T followed by an increase in field to $6$~T results in a line shape with a
negative $\alpha $ (see Fig.2), whereas field-cooling in $7$~T followed by a
reduction of the field to $6$~T results in a positive $\alpha $ (not shown).

\begin{figure}[tbp]
\includegraphics[scale=.3]{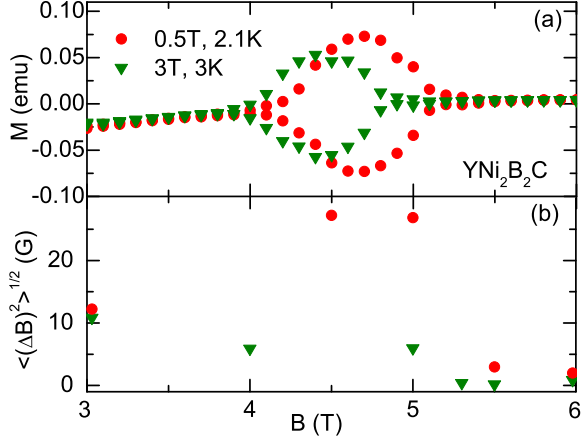}
\caption{(color online) (a)Field dependence of the longitudinal magnetic
moment of YNi$_{2}$B$_{2}$C near the PE region obtained by SQUID
magnetization measurements for the indicated cooldown conditions at 2.1 and
3 K. (b)Field dependence of the $\protect\mu $SR field-distribution width
calculated from the data obtained at 2.1 and 3 K. }
\label{Fig.4}
\end{figure}

Finally, our isothermal SQUID magnetization measurements, performed on both
LuNi$_{2}$B$_{2}$C and YNi$_{2}$B$_{2}$C samples, are shown in Figs.\ \ref%
{Fig.3} and \ref{Fig.4}, respectively. The magnetization data around the PE
regions are plotted together with the width of the TF-$\mu $SR line shapes.
The sharp maximum in the $\mu $SR line-width, shown for both LuNi$_{2}$B$%
_{2} $C (Fig.\ \ref{Fig.3}) and YNi$_{2}$B$_{2}$C (Fig.\ \ref{Fig.4}),
closely follows the PE in the corresponding magnetization curve. However,
the magnitude and field range of the PE strongly depend on temperature.
Consequently, the significant difference between TF-$\mu $SR line widths
observed after field-cooling in $0.5$~T to $2.1$~K and after field-cooling
in $3$~T to $3$~K, are primarily due to temperature.

Summarizing our experimental results, significant line broadening and
reversed skewness have been observed in $\mu $SR measurements, both
correlated with the observation of a PE and an enhanced damping of the dHvA
oscillations. The origin of this remarkable correlation is most probably due
to enhanced pinning-induced vortex-lattice disorder. However, the enhanced
field inhomogeneity observed in the PE region, approximately 30 G (maximal
field distribution width in the PE) in both LuNi$_{2}$B$_{2}$C\emph{\ }and
YNi$_{2}$B$_{2}$C (Figs.\ref{Fig.3},\ref{Fig.4}), seems to be too small to
attribute the additional damping of the dHvA amplitude to further broadening
of the Landau levels by magnetic-field inhomogeneity.

The following analysis establishes this conclusion: Imagine a charged
quasiparticle (with an effective mass $m^{\ast}$) moving freely in a
two-dimensional spatial field profile $B(\mathbf{r})$, consisting of a large
uniform part $B_0$, plus a small nonuniform part $B_1(\mathbf{r}) \equiv B(%
\mathbf{r}) -B_0$. The distribution of $B_1(\mathbf{r})$ is assumed to be
completely random, so that its ensemble average is $\left\langle B_1(\mathbf{%
r})\right\rangle =0$. The Landau-level width, $\pi /\tau _R$, at the Fermi
energy, $E_F$, corresponding to the unperturbed cyclotron frequency, $%
\omega_c=eB_0/m^\ast c$, due to the inhomogeneous broadening, is given in
the semiclassical limit, $n_F=E_F/\hbar \omega_c \gg 1$, by \cite{Aronov95}: 
$\frac{\pi }{\tau _{R}\omega _{c}}=\left( \frac{m^{\ast }E_{F}}{\hbar ^{2}}%
\right) ^{1/2}\frac{\sqrt{\left\langle b^{2}\right\rangle }}{B_{0}}, $ where 
$\left\langle b^{2}\right\rangle =\int \left\langle B_1(\mathbf{r}) B_1(%
\mathbf{r}^\prime)\right\rangle d^2r^\prime$. For a random vortex
distribution near $H_{c2}$ the correlation length is of the order of the
minimal (magnetic) length, $a_{B_{0}}\equiv \sqrt{c\hbar /eB_0}$, and $%
\left\langle b^2\right\rangle \simeq B_1^2a_{B_{0}}^2$, with $%
B_1^2=\left\langle B_1^2(\mathbf{r}) \right\rangle $, so that: {\small 
\begin{equation}
\frac{\pi}{\tau_{R}\omega_{c}}=n_{F}^{1/2}\left( \frac{B_{1}}{B_{0}}\right).
\label{AronovWidth}
\end{equation}
}

The damping of the dHvA amplitude associated with the Landau-level
broadening described above can be estimated from the Dingle factor: $%
R_{D}=\exp \left( -\frac{\pi }{\tau _{R}\omega _{c}}\right) $. This may be
compared with the extra damping factor due to direct scattering of a
quasiparticle by the pair potential in the random vortex distribution limit 
\cite{Maniv01,Maki91}, which takes the form: {\small 
\begin{equation}
R_{M}=\exp \left( -\frac{\pi }{\tau _{M}\omega _{c}}\right) ,\quad \frac{\pi 
}{\tau _{M}\omega _{c}}=\pi ^{3/2}\frac{\widetilde{\Delta _{0}}^{2}}{%
n_{F}^{1/2}},  \label{Maki}
\end{equation}%
} where $\widetilde{\Delta _{0}}\equiv \Delta _{0}/\hbar \omega _{c}$, and $%
\Delta _{0}$ is the self-consistent Ginzburg--Landau expression for the
amplitude of the SC order parameter. At the PE field position: $%
B=B_{PE}\lesssim H_{c2}$, i.e., $B_{PE}\simeq 6$ T, with $H_{c2}\simeq 7$ T
(at about 3~K), we find $\widetilde{\Delta _{0}}^{2}\approx
0.36n_{F}(1-B_{PE}/H_{c2})$, so that combining Eq.\ (\ref{AronovWidth}) with
Eq.\ (\ref{Maki}) one has: {\small 
\begin{equation}
\frac{\tau _{M}}{\tau _{R}}\cong \frac{1}{2\left( 1-B_{PE}/H_{c2}\right) }%
\left( \frac{B_{1}}{H_{c2}}\right) =2\times 10^{-3}.  \label{thauM/thauR}
\end{equation}%
}

The Dingle plot obtained from Eq.\ (\ref{Maki}), modified by thermal
fluctuations \cite{Maniv06}, for reasonable values of the adjustable
parameters $\Delta _{0}$ and $H_{c2}$ \cite{Bergk09}, is shown in Fig.\ \ref%
{Fig.1} to agree well with the experimental Dingle plot in the PE region.
Thus, Eq.(\ref{thauM/thauR}) implies that the additional damping rate
associated with the enhanced field inhomogeneity observed in the PE region
is at least two orders of magnitude smaller than the observed damping rate
shown in Fig.\ \ref{Fig.1}.

The above estimate indicates that in order to reasonably account for the
striking drop observed in the dHvA amplitude at the PE region one should
invoke the direct influence of the SC pair potential on the fermionic
quasiparticles under quantizing magnetic field, rather than the indirect
effect through the magnetic-field inhomogeneity induced by the SC currents.
Within this interpretation the observed recovery of the dHvA signal below
the PE region, shown in Fig.\ \ref{Fig.1}, is explained by the reduced
additional damping of the dHvA amplitude in the Abrikosov vortex lattice 
\cite{Maniv01,Kita02}, as compared to that predicted for the random vortex
distribution \cite{Maniv01}.

This research was supported by the Israel Science Foundation Grant No.\
425/07, by Posnansky Research fund in superconductivity, and by EuroMagNET
under the EU contract No.\ 228043. J. E. Sonier acknowledges support from
the Natural Sciences and Engineering Research Council of Canada. Work at the
Ames Laboratory was supported by the Department of Energy, Basic Energy
Sciences under Contract No. DE-AC02-07CH11358.

This article is dedicated to the memory of Tal Maniv, our beloved son and
grandson.

\end{document}